# **Title**: SpaceTx: A Roadmap for Benchmarking Spatial Transcriptomics Exploration of the Brain

**Authors and affiliations**: Brian Long[1], Jeremy Miller[1], The SpaceTx Consortium[2]

1. Allen Institute for Brain Science, Seattle, WA, USA
2. Consortium authors:

| Name | Affiliation | ORCID |
|---|---|---|
| Boudewijn Lelieveldt | Leiden University Medical Center; Delft University of Technology | 0000-0001-8269-7603 |
| Charles Comiter | Massachusetts Institute of Technology; Broad Institute of MIT and Harvard; Massachusetts General Hospital / Harvard Medical School | 0000-0003-4201-1739 |
| Christoffer Mattsson Langseth | Science for Life Laboratory, Department of Biochemistry and Biophysics, Stockholm University, 171 65 Solna, Sweden | 0000-0003-2230-8594 |
| Daniel Gyllborg | Science for Life Laboratory, Department of Biochemistry and Biophysics, Stockholm University, 171 65 Solna, Sweden | 0000-0002-1429-6426 |
| Ed Lein | Allen Institute for Brain Science, Seattle, WA, USA | 0000-0001-9012-6552 |
| Eeshit Dhaval Vaishnav | Massachusetts Institute of Technology; Broad Institute of MIT and Harvard; Sequome, Inc. | 0000-0003-3720-8051 |
| Jennie Close | Allen Institute for Brain Science | 0000-0001-9234-5855 |
| Jeongbin Park | School of Biomedical Convergence Engineering, Pusan National University, Korea | 0000-0002-9064-4912 |
| Jeroen Eggermont | Department of Radiology, Leiden University Medical Center, Leiden, The Netherlands | 0000-0003-0361-9469 |
| Kun Zhang | Department of Bioengineering, University of California at San Diego, La Jolla, CA, USA | 0000-0002-7596-5224 |
| Markus M. Hilscher | Science for Life Laboratory, Department of Biochemistry and Biophysics, Stockholm University, 171 65 Solna, Sweden | 0000-0001-7782-0830 |
| Mats Nilsson | Science for Life Laboratory, Department of Biochemistry and Biophysics, Stockholm University, 171 65 Solna, Sweden | 0000-0001-9985-0387 |
| Mehrtash Babadi | Data Sciences Platform, Broad Institute of MIT and Harvard, 415 Main St, Cambridge, MA 02142 | 0000-0003-1829-8397 |
| Philip R Nicovich | Cajal Neuroscience, Seattle, WA | 0000-0002-8517-4469 |
| Renee Zhang | J Craig Venter Institute, La Jolla, CA, USA | 0000-0003-2707-5881 |
| Richard H. Scheuermann | J Craig Venter Inst., La Jolla, CA, USA; Department of Pathology, University of California, San Diego, CA, USA; Division of Vaccine Discovery, La Jolla Inst. for Immunology, La Jolla, CA, USA | 0000-0003-1355-892X |
| Sten Linnarsson | Karolinska Institute, Department of Medical Biochemistry and Biophysics, Stockholm, Sweden | 0000-0002-3491-3444 |
| Tamim Abdelaal | Leiden University Medical Center; Delft University of Technology | 0000-0003-2719-5736 |
| Trygve E. Bakken | Allen Institute for Brain Science | 0000-0003-3373-7386 |
| SpaceJam Hackathon Participants | https://spacetx.github.io/people.html | |

## Abstract

Mapping spatial distributions of transcriptomic cell types is essential to understanding the brain, with its exceptional cellular heterogeneity and the functional significance of its spatial organization. Spatial transcriptomics techniques are hoped to accomplish these measurements, but each method uses different experimental and computational protocols, with different trade-offs and optimizations. In 2017, the SpaceTx Consortium was formed to compare these methods and determine their suitability for large-scale spatial transcriptomic atlases. SpaceTx work included progress in tissue processing, taxonomy development, gene selection, image processing and data standardization, cell segmentation, cell type assignments, and visualization. Although the landscape of experimental methods has changed dramatically since the beginning of SpaceTx, the need for quantitative and detailed benchmarking of spatial transcriptomics methods in the brain is still unmet. Here, we summarize the work of SpaceTx and highlight outstanding challenges as spatial transcriptomics grows into a mature field. We also discuss how our progress provides a roadmap for benchmarking spatial transcriptomics methods in the future. Data and analyses from this consortium, along with code and methods are publicly available at https://spacetx.github.io/.

## Introduction to the SpaceTx consortium

Determining spatial distributions of molecularly defined cell types is a core element of the Human Cell Atlas[1], the Brain Initiative Cell Census Network (BICCN[2]) and other efforts to localize transcriptomic cell types at the single cell level in tissue. Single cell transcriptomics methods continue to generate large amounts of data and insight into molecular cell types, albeit with spatial resolution limited by dissections and cell type complexity. In this context, spatial transcriptomics methods are a critical bridge to understanding tissue organization, especially in the brain with its exceptional cellular heterogeneity and functional spatial organization[3,4]. Prior to the outset of SpaceTx in 2017, a large number of such methods were already published[5–17] which each use different strategies for tissue processing, imaging, data processing, analysis and visualization. In 2017 and now, making comparisons across methods and integrating studies derived from diverse methods is a substantial challenge. The SpaceTx Consortium (SpaceTx) was formed with funding from the Chan Zuckerberg Initiative (CZI) to enable a direct comparison of methods by applying these methods to a common experiment: measuring the spatial distribution of previously defined transcriptomic cell types in the mouse primary visual cortex and human temporal cortex. Beyond the focused work in mouse tissue, progress in the much more challenging human cortex was stimulated by making human cortical tissue available for consortium members. This consortium included leading spatial transcriptomics labs, along with the Allen Institute for consortium management and tissue distribution, and support from CZI for software development and data access. We distributed the consortium efforts across six distinct areas: 1. Tissue distribution and laboratory techniques; 2. Reference cell type definition through clustering of RNA-Seq data; 3. Gene panel selection and probe design; 4. Data acquisition; 5. image analysis and data formats; 6. Cell segmentation optimization 7. cell type assignment and meta-analysis; 8. data visualization arising later. These activities illustrated in **Figure 1** and described in more detail below, highlighting our progress and the remaining challenges in each area. Data and analyses from this consortium are publicly available at https://spacetx.github.io.

At the outset of this project, the spatial transcriptomics methods were cutting-edge technologies that had been applied in only a small number of successful experiments, resulting in zero to a few publications each and with no publications outside the originator labs. Experimental protocols, raw data, processing code and analyzed data were difficult to access and not standardized. Since then, spatial transcriptomics methods have improved and become widely available through commercial products, greatly reducing the technical barriers to spatial transcriptomics data generation. However, the need for benchmarks, quality metrics and neutral assessments clearly remains, as users need to select the method that best suits their particular needs in terms of resolution, multiplexing, sensitivity and throughput. In the sections below, we highlight how the work of SpaceTx provides a roadmap for a generating and analyzing benchmark spatial transcriptomics data in the context of large-scale cell type atlases in the brain.

**Figure 1:** SpaceTX consortium workflow for cell type classification in adult mouse and human cortex

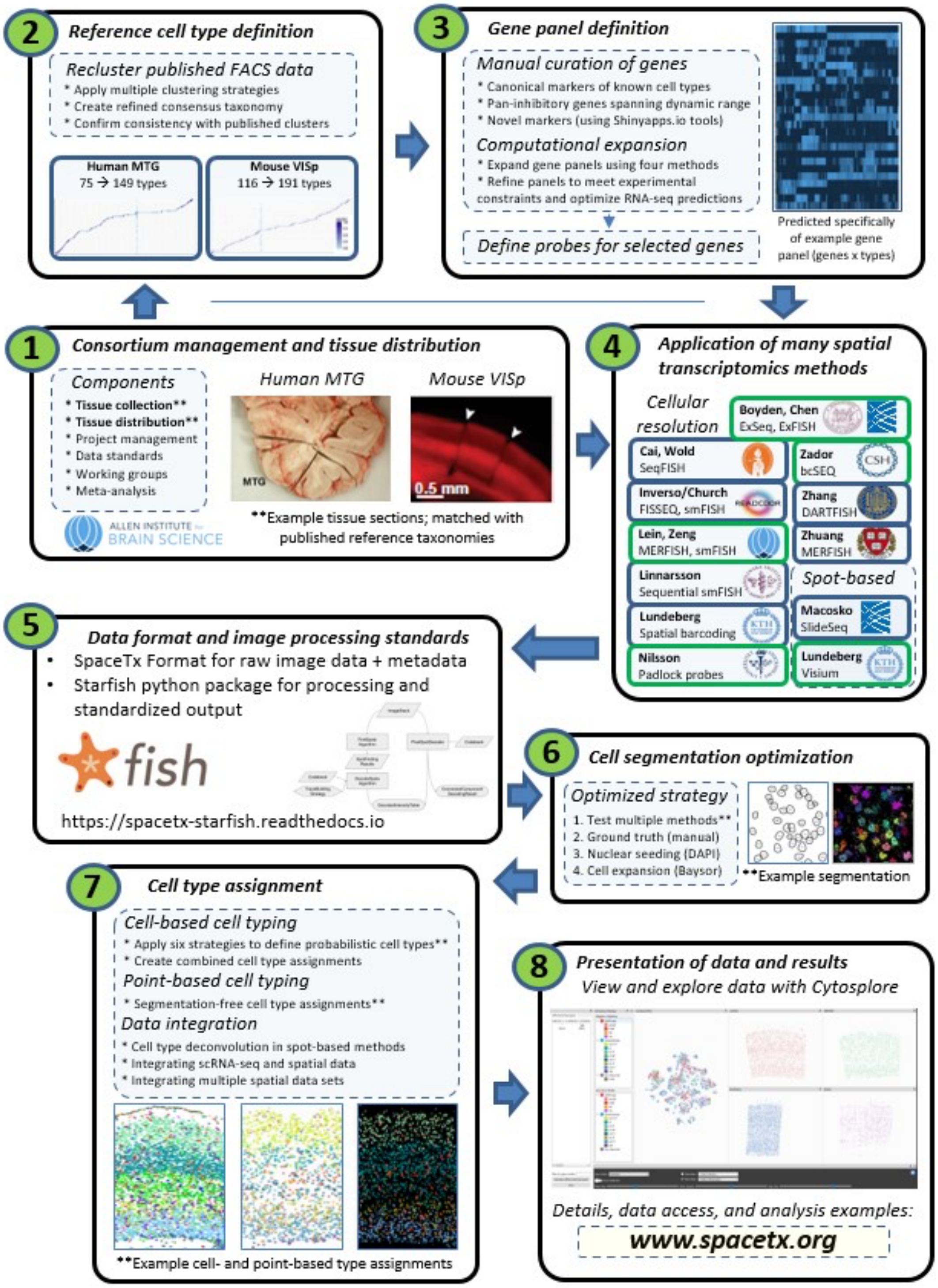

## SpaceTx Consortium experimental methods

The consortium began with 13 labs from 11 universities and institutes who developed 10 distinct spatial transcriptomic methods, with the goal of applying all methods to mouse primary visual cortex and human middle temporal gyrus (MTG). Barcoded methods such as MERFISH [6,12–14] , ISS [9] and BaristaSeq [18] allow for larger gene count and more efficient use of multiround imaging, while non-barcoded methods like multiround smFISH or "osmFISH" [5,19,20] can include more highly-expressed genes without overcrowding probes and are more forgiving for probe loss or imaging errors. Some of these methods include enzymatic amplification: Baristaseq [18], DARTFISH [21] and ISS [9] increase the fluorescence signal allowing faster image acquisition. ExSeq [22] expands the tissue to allow imaging with resolution beyond the diffraction limit. Finally, spot-based sequencing methods like Visuim [8,17] and Slide-seq [23] allow a genome-wide assessment of gene expression, at the expense of spatial resolution as they typically include multiple cells per spot. In principle, each method provides a distinct combination of these and other advantages and drawbacks, but they all should allow spatial localization of cell types defined using single cell or nucleus RNA-seq (scRNA-seq). **Table 1** includes descriptions, specifications, and associated references of each method as applied to mouse VISp and included in the final cell type mapping assessment in SpaceTx.

| method name | readout mechanism | enzymatic amplification | reference | n_genes (SpaceTx) | area imaged (SpaceTx) | tissue thickness (SpaceTx) | detection sensitivity | average total counts per cell (SpaceTx) | time per experiment | Min. expression (FPKM) | Max. expression (FPKM) | Min. gene length (bp) |
|---|---|---|---|---|---|---|---|---|---|---|---|---|
| Allen multiround smFISH | non-barcoded | no | 20 | 22 | ~3 $mm^2$ | 10 m | high | 225 | 1 week | 10 | 300 | 960 |
| Allen MERFISH | barcoded | no | | 254 | ~ 1.9 $mm^2$ | 10 µm | high | 919 | 24 hours imaging, 4 days sample prep | 0 | 25 | 1000 |
| hybridization-based in situ sequencing | sequencing-based | yes | 24 | human: 120, mouse: 119 | human: 26.6 $mm^2$, mouse: 53.6 $mm^2$ | 10 µm | low | human: 38.6, mouse: 30 | <1 week | 100 | N/A | 30 |
| ExSeq | sequencing-based | yes | 22 | mouse: 42 | mouse: ~1 $mm^2$ | 20 µm | medium | mouse: 177 (for QC pass cells) | 4 days imaging; 1 wk sample prep | 0 | 250 | 500 |
| BaristaSeq | sequencing-based | yes | 18 | 76 | .75 x 1.2 mm | | low | 70 | | 100 | N/A | 800 |
| Visium | spatial barcoding, bulk sequencing | no readout imaging | 17 | Full Transcriptome | 6.5mm x 6.5mm | 10 µm | n/a | n/a | ~1 day + sequencing time | N/A | N/A | N/A |

**Table 1: SpaceTx data and experimental constraints**. Sensitivity is based on reported counts per cell as: high: same order of magnitude (OoM) as smFISH, medium: ~1 OoM lower than smFISH, low: more than 1 OoM below smFISH. Constraints on gene expression and gene length were submitted by SpaceTx consortium members in 2017.

## SpaceTx tissue distribution and processing

When SpaceTx convened, tissue processing and wet-lab protocols for spatial transcriptomics varied across labs and were often difficult to reproduce. One critical aspect of data standardization and benchmarking is ensuring that the starting material that each group has to work with is as similar as possible (**Figure 1**) and so the Allen Institute for Brain Science supplied human brain tissue and specified conditions for tissue collection from mouse primary visual cortex. For human, the Allen Institute for Brain Science collected slabs of neurosurgically resected tissues from 3 cases with sufficient tissue in MTG, divided these into 300 µm thick blocks with a tissue slicer, flash froze these blocks fresh, and shipped these adjacent sections for groups to section and process optimally for their methods. For mouse, each group collected tissue slabs from adult mice (P56 - P60), dissected VISp on their own, and shared block face images with the Allen Institute for anatomical delineation and confirmation. SpaceTx also worked on improving some of the tissue processing steps critical to multiple methods, including stripping/hybridization methods, conjugated labels, and membrane labeling. Lipofuscin, an autofluorescent protein that aggregates in brain tissue with age[25], presented a particularly robust challenge when dealing with human tissues. While no single lipofuscin removal strategy worked for all groups, a combination of chemical treatment (e.g., with TrueBlack) and photobleaching reduced autofluorescence to a manageable level in some cases (**Supplemental Figure 1**). Despite these challenges, making human cortical tissue available for investigation resulted in two publications that were the first papers detailing the spatial organization of human cortical cell types using cellular-resolution spatial transcriptomics[26,27]. Finally, several of the protocols for tissue processing and other SpaceTx activities were deposited in Protocols.io for method reproducibility[28].

**Benchmarking Recommendation 1. Tissue Distribution and Processing:**

- Define specific samples for measurements, a well-defined anatomical region where tissue collection can be standardized.
- Define common tissue sources and processing conditions for specimens distributed for benchmarking.
- Document protocols and details about reagents (product numbers, lot numbers, etc.).
- Reagent Reporting Standards
    - Include target sequence information for each transcript targeted
    - Include part numbers for reagents and the acquisition system used for data collection.

## VISp and MTG reference taxonomy development

The recent explosion of single cell omics papers and associated taxonomies has highlighted the importance of standardizing cell type taxonomies and nomenclature. At the outset of SpaceTx, we built reference datasets for mouse VISp and human MTG using public data and associated analyses as a starting point [4,29]. There were two primary criteria for building novel reference taxonomies. First, cells from reference taxonomies were assigned to types using less stringent criteria than in the initial publications, to allow a more precise matching of spatial data to reference data. In particular, a clustering strategy was chosen to subdivide cell types with strong spatial gradients, such as human L2/3 IT neurons and Pvalb interneurons (**Supplemental Figure 2**). Second, since there was not yet an established, reliable computational clustering strategy, a method was selected that relied on the output from multiple clustering algorithms. Specifically, the clusterExperiment R package[30] was used to create a consensus reference taxonomy based on results of four clustering methods (SIMLR [31], PAGODA [32] scrattch.hicat [29], and proMMT [33]) run with different parameter settings, which resulted in 149 human MTG and 191 mouse VISp clusters total. Notably, this represents approximately twice the number of clusters that have been published based on these datasets and therefore represents an exceptionally high-resolution taxonomy. In the intervening years, single-cell RNASeq datasets have grown in scale and analysis methods have grown in sophistication. Defining reference taxonomies on large datasets spanning millions of cells across multiple tissue regions is its own challenge, one made more feasible through corroboration with spatial methods. Despite these challenges, a defined taxonomy provides expectations for gene expression measurements from spatial transcriptomics and context for interpretation of mapped cell types.

**Benchmarking Recommendation 2. Define a Target Cell Type Taxonomy:**

- Design benchmarking experiments to sample and distinguish a pre-defined and widely-accepted cell type taxonomy within the chosen anatomical area(s).

## Gene panel selection

Prior to the SpaceTx consortium, selection of gene panels for spatial transcriptomics was done using a combination of computational and manual selection, frequently without a direct reference to a specific target taxonomy. With our experimental goal of mapping spatial transcriptomics cells to defined clusters, the need for a principled approach to gene panel selection was clear. Using each reference scRNAseq taxonomy as a target, participating groups developed gene sets optimized for each spatial transcriptomics. To the extent possible, gene panels were designed to (1) distinguish all consensus cell types, (2) include some marker genes with known expression patterns, (3) test the dynamic range of each experimental method, (4) meet the predicted constraints of each experimental method, (5) maximize the overlap between experimental methods, (6) build in redundancy to allow for probe failures, and (7) include genes selected using several computational strategies to maximize the chances for success (**Figure 1**). Public-facing analysis and visualization tools were created to aid in selection of high-confidence marker genes (Sifter[34]), and expansion of gene panels and assessment of accuracy (SWIMS[35]). Panels were selected following the design principles laid out above, starting with a manually selected initial set of ~20 genes and expanded using a combination of four computational strategies (greedy algorithm used in mfishtools[36] , probability model[33], NS-Forest[37], and random forest) which were rank ordered based on information content. In total, four ordered sets of marker genes were generated for each species, with each experimental method assigned to a specific set based on the number of genes to run and the constraints on selected genes[38,39]. Finally, genomic locations within each gene with assigned reads in the snRNA-seq data were identified and provided to each experimental group to ensure probes were designed for gene isoforms present in the cortex.

The scale of spatial transcriptomics experiments has greatly expanded since the end of SpaceTx, with a few groups producing data sets sampling the entire mouse brain (BICCN, unpublished studies in progress). With substantially more cell type complexity than the mouse primary visual cortex alone, selecting a successful gene panel to cover the entire mouse brain will require specialized analysis beyond the approaches used here. Researchers have recently made innovations in this area (geneBasis[40], PROPOSE[41], activeSVM[42], Spropos[43] but so far, there has been little direct comparison of gene panel selection methods in the context of experimental spatial data. SpaceTx experiments were unable to “close the loop” linking cell type mapping performance back to gene panel design, but with current commercial methods capable of probing hundreds of genes in an experiment, gene panel selection methods could be compared in a single experiment by performing cell type mapping using subsets of genes chosen by different gene selection methods. Results from this kind of experiment could be used to improve computational methods for gene panel selection.

Spatial methods have limitations on the range of gene expression that can be probed in an experiment (see Table 1). Detection sensitivity sets the low end of gene expression, whereas imaging and codebook properties impact the high end of gene expression. When imaged mRNA molecules are denser than can be resolved by the optical resolution of the microscope, readout decoding can suffer in barcoded methods such as MERFISH, and molecular quantification of non-barcoded methods such as osmFISH can become inaccurate. These experimental constraints differ across experimental methods and SpaceTx labs addressed this by supplying upper and lower limits of FPKM from bulk RNASeq in the same tissue. This was a helpful starting point, but it doesn’t take into account the gene panel composition and expected total number of mRNA probed in each cell type, which is the source of optical crowding-based upper limits on gene expression. Future gene panel selection methods should incorporate these experimental limitations and optimize gene panels to avoid artifacts from optical crowding in a cell type-specific manner.

**Benchmarking Recommendations 3. Gene Panel Selection:**

- To the extent possible, identify genes that can perform well for cell type discrimination while also satisfying experimental constraints of all methods
- Use one or more experimental methods that probe large numbers of genes to combine panels from different selection algorithms and allow comparisons within a single experiment

## Image processing and data standardization

During the initial organization of SpaceTx, the similarities of raw data across methods presented an opportunity for standardized data formats and image processing. Imaging-based methods of spatial transcriptomics described here all utilize a common procedure for microscopy data collection: multiround, multichannel imaging in a spatially tiled acquisition. Between rounds of imaging, different readout chemistry is applied, allowing each round and channel to reflect a different readout bit of the underlying encoding scheme. In the case of multiround smFISH, each channel in each round shows the fluorescent oligos bound to one species of mRNA. In other methods, the relationship between images and targeted mRNA is more complex, with each image in a multiplexed method such as MERFISH containing all mRNA locations from ~40 different targeted mRNA species. The commonalities across methods inspired two major open-source software efforts in the SpaceTx consortium by CZI engineers and computational biologists in concert with consortium scientists: First, SpaceTx Format[44]: This raw image data format is built around the concept of a tiled, multiround, multichannel experiment and is readable from cloud locations (AWS S3) or a local filesystem. Extensive documentation and examples are available to help users convert existing data and metadata into SpaceTx format. SpaceTx format has some limitations that stem from the tiled acquisition model, but it could be extended to accommodate alternative acquisition schemes that do not create tiled images. To our knowledge, it remains the only fully defined format for raw data from imaging transcriptomics experiments. Starfish[45]: This python package for end-to-end processing of spatial transcriptomics data includes modules for spotfinding, registration, decoding and segmentation that can be mixed and matched across multiple spatial transcriptomics modalities and decoding schemes. Development of starfish slowed during the Covid-19 pandemic, but has recently been more active, with community members contributing new decoding methods in 2022.

Whether processed with starfish or by using in-house pipelines, SpaceTx-contributed datasets include mRNA locations, segmented cell locations and total counts and a DAPI image of the assayed area: the primary visual cortex in a coronal section of the mouse brain.

**Benchmarking Recommendations 4. Image Processing and Data Standardization:**

- **Processed Data:**
  - Standardize processed data outputs to facilitate downstream processing and comparisons (See Appendix 1)
- **Raw Image Data:**
  - For benchmarking of experimental methods, we recommend storage of at least some raw data for every experiment. This will allow community members to verify decoding results and develop improved algorithms.
  - As data volumes increase, costs for raw data transfer and storage may not be feasible for large-scale experiments.
  - (For further details, see Appendix 1)

## Cell segmentation

After mRNA identities and locations are extracted from raw images, the next step is typically cell segmentation (with a few exceptions noted below). The product of cell segmentation is a cell-by-gene table with the counts of each gene assigned to cells, along with the spatial location and other parameters such as the cell's area or volume and potentially its anatomical location. The SpaceTx consortium tested two general approaches for cell segmentation during our "SpaceJam" hackathons. First are cell segmentation methods without transcriptomic priors, which utilized DAPI images that provide nuclear stains as input. Three such methods were tested: a feature-based approach involving foreground segmentation and consecutive splitting into separate nuclei, a learning-based approach, and an example case of expert manual annotation of DAPI nuclei for use as ground truth. Overall, these methods produced relatively comparable results, except in cases with overlapping cells or cells reaching out of the plane of sectioning. The second approach is augmented cell segmentation, which utilizes spot by gene matrices from the spatial transcriptomics experiments themselves for both cell segmentation and cell type assignment. Two such methods were tested (pciSeq[46] and Baysor[47]), which had moderately compatible results in most cases. These methods accurately segment the cell soma borders of adjacent cells of distinct types, while DAPI-based approaches excel at segmenting nuclei regardless of type. For this project, cells were segmented using a combination approach, where DAPI images were segmented using the feature-based approach when available, and these results were used to seed Baysor for final cell segmentation. Called cells were then quality filtered to exclude obvious artifacts from consideration, and the remaining cells were used for downstream cell type assignments.

This segmentation pipeline was adopted through an organic process of iteration and comparison across methods, but the general approach (nuclear segmentation followed by probabilistic assignment of mRNA to cells) has potential to be widely useful. A limitation of augmented segmentation methods is that they are sensitive to the genes and cell types present in the data. For example, one challenge in brain tissue is correctly separating neuronal cells from closely-opposed nonneuronal cells; in this situation, augmented cell segmentation methods would likely produce different results depending on the presence of neuronally-expressed vs non-neuronally expressed genes in the panel. Furthermore, as datasets increase in size and number of genes, augmented segmentation methods may face computational challenges with increased complexity. Recent efforts in applying machine learning methods to general bio-image cell segmentation have been quite successful, including stardist[48] for nuclear segmentation and cellpose[49] for dissociated cells. In addition to cell segmentation, we applied a segmentation-free method (SSAM[50]) to the SpaceTx datasets to characterized transcriptomic diversity in the tissue without creating a cell by gene table *per se*. Recent work outside the consortium also includes another augmented approach (ClusterMap) that approaches spatial transcriptomics analysis as a multi-scale clustering problem, using clusters of mRNA locations to define cells and clusters of cells to define tissue domains[51]. In this way it is very similar to the SSAM approach, but clustermap is based on the mRNA locations directly, whereas SSAM utilizes kernel density estimation to spatially sample mRNA densities and cell identities.

For atlas-scale spatial transcriptomics datasets, segmentation errors can be difficult to identify or correct among millions of cells across multiple tissue areas and many cell types. In some cases, segmentation errors can be identified after cell type mapping if undersegemented cells coexpress mutually exclusive marker genes, but systematic approaches for segmentation quality assessment remain unexplored.

Further improvements in cell segmentation will be enabled by gold-standard data generated by human annotators. Although this gold-standard data does not need to be at a large scale, it should cover as many species and tissue regions as possible to allow for development of robust segmentation algorithms. This data would also be helpful in developing segmentation quality metrics that could be measured independent of gene panel design or cell taxonomy. Such quality metrics would allow assessment of new datasets and algorithms in the future without further development of gold-standard datasets.

**Benchmarking Recommendations 5. Segmentation:**

- Generate gold-standard segmentation data to encourage development of new segmentation algorithms.
- Develop segmentation quality metrics, ideally metrics that are independent of gene panel design, segmentation strategy or cell type taxonomy.

## Cell type assignment

The scientific goal of SpaceTx was to measure the spatial locations of transcriptomically-defined cell types from multiple experimental modalities, so a primary focus was assigning each segmented cell to a cell type in the reference taxonomy. Importantly, this was a consequence of our experimental design: by focusing on reference cell types from scRNASeq, we could treat cell type assignment as a modular process that could be tested across data from multiple experimental methods. Instead of treating a single assignment method as a gold standard, we focused on comparing and combining cell type assignment methods applied to our SpaceTx experimental data. Cell type assignment efforts are described in detail in a separate manuscript[52], and briefly summarized here: For each experimental method, up to six distinct strategies were used to assign cells to cell types from the taxonomy defined above [29,36,46,53–55], and the results were combined in two different ways to produce final type calls and associated confidence scores (**Figure 1**) While most methods defined cells based on existing cell-by-gene tables, some used other starting points or included multiple data modalities. Three methods worked straight from the spot by gene matrix: both Baysor[47] and pciSeq[46] segment cell and assign cell assignments iteratively, thereby optimizing both steps, while SSAM[50] uses kernel density estimation to assign cell types to each position in the image without the need for segmentation at all.

Other cell type assignment methods include information from scRNA-seq directly into spatial transcriptomic results in order to impute gene expression values, either for each method separately (Tangram)[55] or by including all methods at once (SpaGE)[56]. Finally, even spot-based methods such as Visium[17] and Slide-seq[23] can be analyzed to estimate cell type densities, despite their lower spatial resolution. The problem of cross-modality cell type assignment has received attention in the single-cell 'omics field as researchers seek to create multimodal reference taxonomies (e.g. scArches[57]). In all of the SpaceTx data, the resulting cell type assignments generally matched expected spatial distributions, with glutamatergic neurons localized to the expected cortical layer and GABAergic neurons showing spatial localization and density consistent with published reports. However, the details varied substantially in the final output not only when comparing tissue sections derived from different spatial transcriptomics methods, but also when considering results from different analysis methods applied to the same segmented tissue section. This variability in the details highlights the need for continued work in this area, including comparison of cell type assignment approaches to *de novo* clustering, quantification of the quality of cell type assignment and novel approaches for cell type assignment.

**Benchmarking Recommendations 6. Cell type assignment:**

- Compare cell type assignment methods across experimental approaches.
- Where possible, use domain knowledge experts and annotated scRNASeq data to compare results to known spatial distributions.

## Visualization of results

Interactive data visualization is an essential part of data exploration for single-cell 'omics data, with tools such as cellxgene[58], Cytosplore[59], Scanpy[60], Loupe browser, and Seurat[61] in widespread use. At the beginning of SpaceTx, there were no purpose-built tools or integrated capabilities for visualizing spatial transcriptomics data, so we developed a visualization tool to display data from this project. For this study, the Cytosplore Viewer, was extended for comparative visualization and analysis of the different spatial protocols. This included (1) application of a combined molecular embedding and associated imputed gene expression values (SpaGE[56]) to visualize cells across multiple protocols at once, (2) comparative simultaneous visualization between multiple spatial data sets, and (3) distinct views of data and metadata for the four protocols included in the comparative analyses (smFISH, MERFISH, BaristaSeq, and ExSeq). When viewing expression values, either the measured and imputed values can be painted on the spatial and tSNE maps. Furthermore, functionality for differential expression analysis between two manual selections was implemented, enabling quick retrieval of differentially expressed genes between regions or cell types. Finally, the individual local maxima and associated gene matrices and cell type assignments based on SSAM Kernel Density Estimate profiles were included in the single protocol visualizations for comparison with results from the segmentation-based consensus cell type assignment methods.

During the course of SpaceTx, tools for viewing cell-by-gene single cell data (cirrocumulus[62], cellxgene[58,63]) have been used to view spatial location as an alternative "embedding" of the data. Additionally, visualization tools specifically for spatial transcriptomics data are being developed, including Vitessce[64], Giotto[65], and Squidpy[66]. In the future, these visualization tools will all benefit from standardized data formats which will allow data to be easily visualized in many tools without format conversion or data wrangling. The SpatialData project[67] is one promising effort in this direction. Another challenge for data visualization developers is how to seamlessly integrate multiple datasets for viewing side-by-side. The Cytosplore SpaceTx viewer accomplished this for a fixed group of datasets and an extendable version of this capability would be helpful for comparing other combinations of datasets in the future.

**Benchmarking Recommendations 7. Visualization:**

- Make benchmarking outputs maximally compatible with community-developed visualization tools

## A Roadmap for Benchmarking Spatial Transcriptomics Beyond SpaceTx

The field of spatial transcriptomics has gone through dramatic growth and change since the beginning of the SpaceTx Project. In the early phases of SpaceTx, data generation was a substantial challenge for many of the experimental groups and the lack of data limited progress on data analysis. Once data was available on a single anatomical area, development on cell type mapping and related analyses were able to proceed more freely. These phases mirror how the whole field has progressed: with this project, commercial data acquisition systems and other publicly-available data (e.g., BICCN[66], Vizgen[68], Visium[69], and HuBMAP[70]), analysis of spatial transcriptomics is now a growing field[17,47,50,55,56,66,71]. SpaceTx advanced and standardized several aspects of spatial transcriptomics experiments and associated analyses, but in the meantime, the scale of the spatial transcriptomics data sets has increased dramatically. Many aspects of spatial transcriptomics data storage and processing scale predictably with increased number of cells or imaging area, but as the scale increases from a single region of tissue to a whole organ (such as mouse visual cortex in comparison to the whole mouse brain), the downstream impacts are less predictable. Scientists exploring larger data sets may encounter dramatically higher transcriptomic complexity, greater variation in spatial organization and even differences in cytoarchitecture and gene expression that impact segmentation quality and cell type mapping within one dataset. This highlights the need for creative approaches to accomplish cutting-edge analysis on atlas-scale datasets, such as universal quality metrics to allow for automated parameter optimization, hierarchical cell type mapping and data-driven spatial subsetting of data before analysis.

With commercially available spatial transcriptomics instruments generating data at unprecedented scales, there is an urgent need for spatial transcriptomics benchmarking and standards. Although the SpaceTx Consortium focused primarily on imaging-based technologies, new forms of sequencing-based approaches are approaching single-cell resolution[refs], some with large acquisition areas (3x5 cm reported)[72]. As these techniques improve sensitivity and resolution, their utility for cell type mapping in the brain should be explored alongside imaging-based methods. As spatial transcriptomic technology development continues, the results of the SpaceTx Consortium and the Benchmarking Recommendations presented here can provide a foundation for high quality spatial transcriptomic measurements and a roadmap for community standards.

Appendix 1:

Data Requirements:

- **Processed data:**
    - **Overview images:** Tissue images of all counterstains such as DAPI, PolyT, membrane labels, etc. and accompanying transformations from image pixels to physical coordinates (microns).
    - **Transcript locations:** All detected transcript locations in physical coordinates (microns) and any associated quality metrics. Segmentation results should also be included here as a column indicting which cell each molecule was assigned to.
    - **Internal controls:** All experiments should include internal negative controls such as the “Blank” codewords used in MERFISH experiments. Any measured detection of these negative controls provides an estimate of the noise floor for transcript detection and can be useful for diagnostics.
    - **Segmentation results**:
        - If segmentation methods produce a spatial definition of each cell, these should be available in an accessible file format. Examples include GeoJSON or GeoPandas-compatible files for polygons.
        - Even if segmentation methods produce a spatial definition for each cell, transcript locations should also be annotated with cell assignment information.

Notes on Raw Image Data:

- Raw image data has critical utility in the context of spatial transcriptomics:
    - Debugging experiments:
        - Determining the impact of tissue autofluorescence such as lipofuscin on processed data and downstream outputs
    - Scientific documentation, traceability and reproducibility
        - There are multiple strategies for decoding multiplexed image data into transcript locations and future methods could improve detection or noise rejection. Without raw data, such reanalysis is impossible.
- Retaining raw data also creates substantial costs:
    - Raw image data spatial transcriptomics experiments is routinely on the TB scale and methods with larger imaging area can easily reach 10s of TB per experiment.
    - This image data must be transferred from the acquisition system in such a way that it can be processed to create experimental outputs, accessed for debugging or reprocessing and also archived for long-term storage if desired.

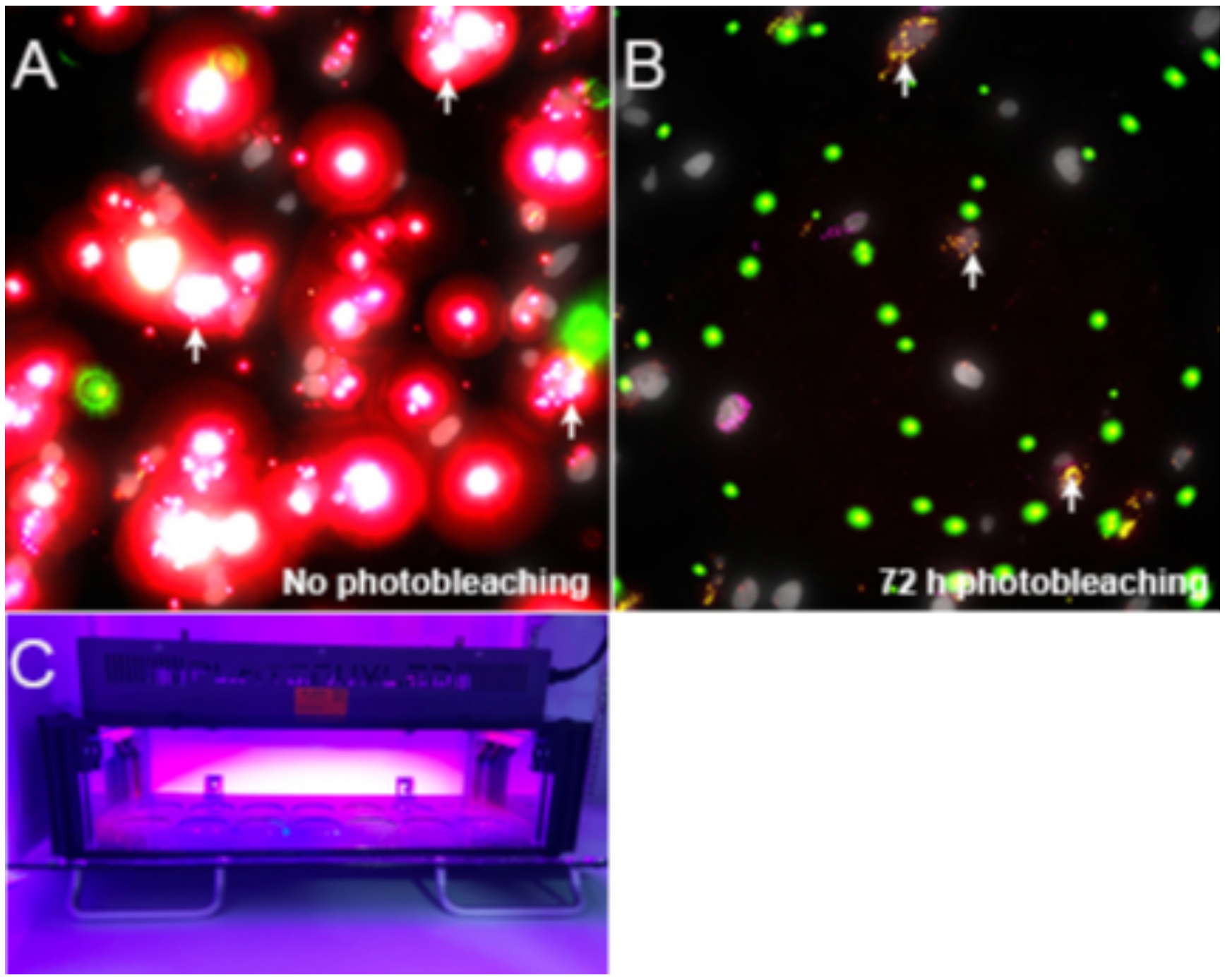

**Supplemental Figure 1: Lipofuscin photobleaching in human tissue.** To reduce the brightness of lipofuscin in adult human brain samples, we photobleach 10um sections in either 2XSSC or 70%Ethanol for up to 72 hours using a 150W LED array. A) Without photobleaching, human brain tissue contains multiple brightly fluorescent lipofuscin granules (indicated by white arrows). B) After 72 h of photobleaching, a section from the same human brain contains reduced lipofuscin levels (example of residual lipofuscin granules indicated with white arrows). mRNA signal can be detected in this tissue: CUX2, red; GAD1, magenta. Green spots are fiducial beads necessary for focus and registration.  C) Photobleaching illustration.

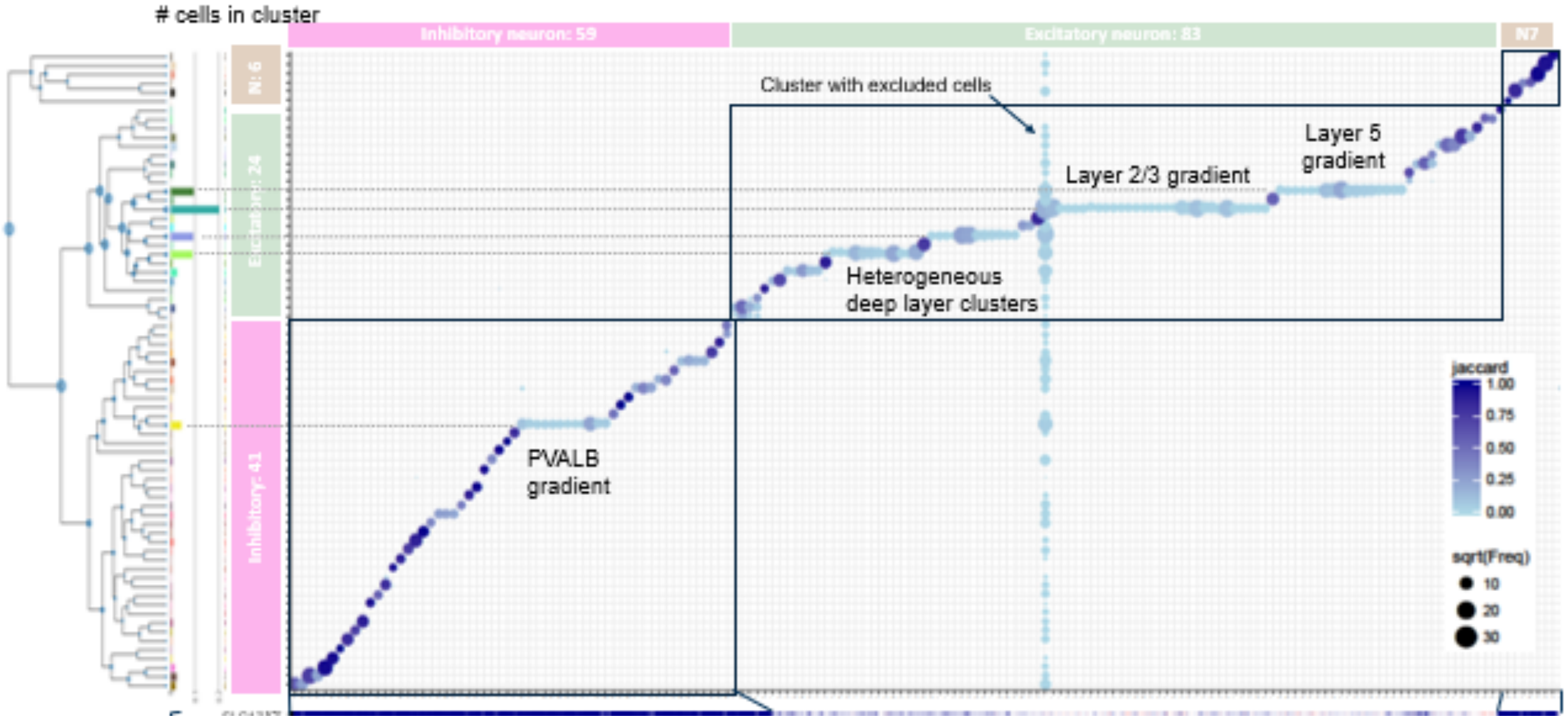

**Supplemental Figure 2: Correspondence between reference cell types and SpaceTx cell types in human MTG.** Horizontal: Cell type dendrogram and counts for cell types defined in[4]. Vertical: Cell type dendrogram and cell counts for cell types defined in SpaceTx. Matrix represents the confusion between these two values, with dot color indicating Jaccard distance between clusters and dot size indicating the number of cells in corresponding pairs of clusters. Overall, there was good agreement between clusters identified in the two studies (e.g., few off-diagonal elements) with a few notable exceptions: (1) more clusters were intentionally identified in SpaceTx (149 vs. 75), (2) types with reported gradients such as L2/3 IT and other excitatory types and PVALB+ interneurons split into dramatically more types, and (3) additional cells were excluded from SpaceTx cell type assignment.

Acknowledgements: We thank all the SpaceTx hackathon participants for their contributions and helpful conversations.

Funding: This work was funded in part by the Chan Zuckerberg Initiative.